# Spectral features of Pb-related color centers in diamond

Sviatoslav Ditalia Tchernij[1,2,3], Emilio Corte[1,2], Tobias Lühmann[4], Paolo Traina[3], Sébastien Pezzagna[4], Ivo Pietro Degiovanni[3,2], Georgios Provatas[5], Ekaterina Moreva[3], Jan Meijer[4], Paolo Olivero[1,2,3], Marco Genovese[3,2], Jacopo Forneris[1,2,3*]

[1] Physics Department, University of Torino, Torino 10125, Italy
[2] Istituto Nazionale di Fisica Nucleare (INFN), sezione di Torino, Torino 10125, Italy
[3] Istituto Nazionale di Ricerca Metrologica (INRiM), Torino 10135, Italy
[4] Applied Quantum Systems, Felix-Bloch Institute for Solid-State Physics, Universität Leipzig, Leipzig 04103, Germany
[5] Laboratory for Ion Beam Interactions, Ruđer Bošković Institute, Zagreb 10000, Croatia

**Abstract**

We report on the systematic characterization of the optical properties of diamond color centers based on Pb impurities. An ensemble photoluminescence analysis of their spectral emission was performed at different excitation wavelengths in the 405-520 nm range and at different temperatures in the 4–300 K range. The series of observed spectral features consist of different emission lines associated with Pb-related defects. Finally, a room-temperature investigation of single-photon emitters under 490.5 nm laser excitation is reported, revealing different spectral signatures with respect to those already reported under 514 nm excitation. This work represents a substantial progress with respect to previous studies on Pb-related color centers, both in the attribution of an articulated series of spectral features and in the understanding of the formation process of this type of defect, thus clarifying the potential of this system for high-impact applications in quantum technologies.

*Corresponding author: jacopo.forneris@unito.it

1. **Introduction**

Diamond color centers have been extensively studied in the last decades as appealing systems for applications in quantum optics and quantum information processing, ranging from the use as single photon emitters to quantum sensing of magnetic and electric fields and of temperature, pressure and strain[1–7]. Despite the increasing interest of these systems among the scientific community, only a fairly limited number of emitters related to impurities incorporated in the diamond lattice has been demonstrated so far[6,8,9], and most of the preliminary practical demonstrations of diamond-based quantum technologies rely on the peculiar spin properties of the negatively-charged nitrogen-vacancy color center (NV)[10–15].

Indeed, even if NV centers present several interesting properties, they are also subject to significant drawbacks, such as broad spectral emission, charge state blinking, relatively low emission rate[16,17], thus prompting for the search of alternative new centers overcoming this limits and/or presenting complementary properties.

Besides the already consolidated color centers related to group-IV impurities such as SiV, GeV, and SnV complexes[6,18–27], two independent works have recently demonstrated the creation of optically-active defects in diamond upon Pb ion implantation and subsequent annealing, and have reported their characterization at the single-photon emitter level[28,29].

The Pb-related defects exhibited appealing emission properties for quantum information processing applications, with room temperature photoluminescence (PL) emission rates exceeding $10^6$ photons s$^{-1}$ [28], an emission mainly concentrated in the zero phonon line (ZPL) due to the low phonon coupling and a large ground-state splitting predicted by numerical simulations[29,30], with potentially disruptive implications towards the achievement of long spin coherence for quantum memories and optical quantum networks[6].

On the other hand, the currently available reports evidenced the presence of several spectral lines of uncertain attribution. Furthermore, discrepancies in the observation of some emission peaks (namely: 520-525 nm doublet, 538 nm doublet, 715 nm line) emerged and were attributed to different experimental conditions such as excitation wavelength and operating temperatures[28,29].

We report here on a systematic investigation of Pb-related centers under varying excitation wavelengths in the 405-520 nm range and temperature between 4 K and 300 K. Our results address and discuss the afore-mentioned discrepancies by providing systematic information on the relative emission intensities of these spectral features as a function of the measurement temperature, thus paving the way to the practical application of these centers. Finally, a comprehensive discussion summarizing the currently available results is presented, including the possible attribution of the observed emission lines and their relation with the theoretical models proposed in the literature.

2. **Experimental**

The experiments were performed on a single-crystal IIa diamond substrate produced by ElementSix. The sample has a "detector grade" quality (i.e. substitutional N and B concentration: $[N_S] < 5$ ppb, $[B_S] < 5$ ppb). Pb-related centers were created upon the implantation of 35 keV $PbO_2^-$ ions at different fluences, namely $5 \times 10^{11}$ cm$^{-2}$ and $2 \times 10^{13}$ cm$^{-2}$. The sample underwent a thermal annealing process at 1200 °C for 2 hours in high vacuum ($<5 \times 10^{-6}$ mbar) to promote the formation of the centers, as discussed in a previous report[28]. It is worth remarking that the implantation of $PbO_2^-$ ions was motivated by the higher extraction efficiency and stability with respect to the technically challenging acceleration of a $Pb^-$ beam[31]. At the same time, a control sample was implanted with 9 MeV $O^{4+}$ ions at $4 \times 10^{13}$ cm$^{-2}$ fluence and annealed under the same experimental conditions, in order to explicitly assess the contribution of optically-active O-related defects to the PL spectra acquired in the considered wavelength range[28]. After annealing, an oxygen plasma cleaning process was performed to ensure the removal of the surface graphitic phases and optically-active contaminants.

The processed substrate was characterized using different PL single-photon sensitive confocal microscopy setups. A cryogenic setup operating based on a liquid helium closed-cycle optical cryostat[32] (100× air objective, 0.85 NA) operating in the 4 – 300 K temperature range was employed for the measurements reported in Sect. 3.1. The latter apparatus was not equipped with nano-positioners, thus preventing the investigation of individual emitters at cryogenic temperatures. For this reason, an additional confocal microscope operating at room temperature (100× air objective, 0.95 NA), equipped with closed-loop nano-positioners was used for the identification of individual emitters (measurements reported in Sect. 3.2). In both setups, the optical excitation was provided by a set of CW laser diodes with emission wavelengths of 405 nm, 490.5 nm and 520 nm. The PL spectra were acquired using a single-grating monochromator (1200 grooves mm$^{-1}$, 600 nm blaze, ~4 nm spectral resolution) that was fiber-coupled to a single-photon avalanche detector (SPAD). The room-temperature setup was equipped with a Hanbury-Brown &Twiss (HBT) interferometer, allowing the assessment of single-photon emitters via the measurement of the second-order autocorrelation function. The HBT interferometer was implemented by connecting two independent SPAD to the output of a multimode fiber-integrated 50:50 beam-splitter. A third confocal microscope (60× oil objective 1.35 N.A.) equipped with a SPAD outcoupled in open air was used to characterize the spectral emission from individual spots under 532 nm laser excitation (measurements reported in Sect. 3.3). In this latter setup, the PL spectra were acquired using a Peltier-cooled CCD array camera with a spectral resolution of ~2 nm, as determined by the first-order Raman peak (572 nm) FWHM width.

# 3 Results

## 3.1 Ensemble-level PL characterization

Following preliminary reports on the formation and the optical properties of Pb-related color centers[28,29], we performed the first extensive PL characterization of this class of defects at an ensemble level ($2\times10^{13}$ cm$^{-2}$ implantation fluence) under different excitation wavelengths in a temperature range of 4–300 K. Fig. 1 shows the PL spectra acquired at 4 K, 77 K and 300 K under 405 nm (Fig. 1a), 490.5nm (Fig. 1b) and 520 nm (Fig. 1c) excitations, respectively. The main results of this set of measurements can be summarized as follows.

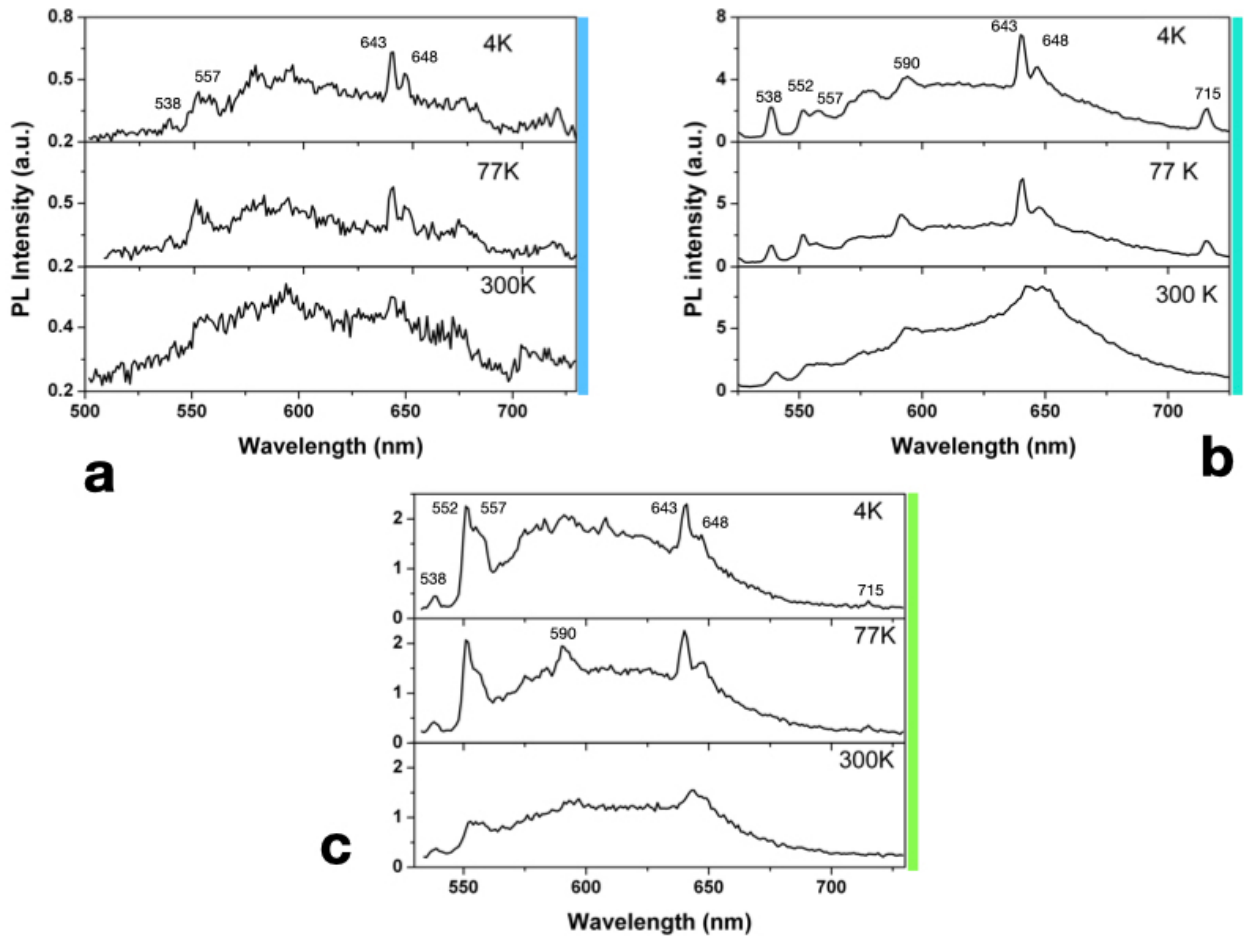

**Figure 1:** PLspectra acquired at different temperatures (4 K, 77 K, 300 K) from the region implanted with PbO$_2^-$ ions at $2\times10^{13}$ cm$^{-2}$ fluence. The reported spectra were acquired under the following optical excitation wavelengths: **a)** 405 nm **b)** 490.5nm **c)** 520 nm.

*405 nm excitation.* The spectra acquired under 405 nm laser excitation (10 mW optical power) are reported in Fig. 1a. All spectra were acquired using along-pass filter with 500 nm cutoff wavelength. The spectra were limited to wavelengths shorter than 720 nm due to the observation of fluorescence at larger wavelengths from the optical components of the confocal microscope. The most evident spectral features, mainly visible at lower temperatures, consist ofa small peak at 538 nm, a band centered at ~557 nm, and a broad band in the 570–670 nm range. Additional sharp peaks at 643 nm and 648 nm are visible only at 4 K and 77 K temperatures. At 300 K temperature, the above-listed peaks convolve into a broader emission band centered at 645 nm. Considering the ~3nm spectral resolution of the apparatus and the spectral blue-shifting at decreasing temperatures (already reported in Ref. [28]), these lines are compatible with the weak emission lines at 640.4 nm and 649.8 nm observed under the same excitation wavelength in Ref. [28].

*490.5nm excitation*. Spectra were acquired at 3.5 mW excitation power, using a laser diode with 490.5 nm emission wavelength. The spectra were acquired using a 500 nm long-pass filter and a (514 ± 15) nm notch filter to prevent the observation of the first-order Raman line (expected at 524.8 nm based on the 1332 cm$^{-1}$ Raman shift of diamond). The main PL emission features(Fig. 1b) consist of peaks located respectively at 538 nm, 552 nm and 557 nm, together with an additional wide band at~590 nm, consistently with what reported in Ref. [28] under 532 nm excitation. The emission lines at 643 nm and 648 nm are also visible under 490.5nm laser excitation. Finally, a PL peak is observed at 715 nm at 4 K and 77 K temperatures, while this spectral line could not be resolved at room temperature. A PL spectrum (Fig. 2) of the O-implanted control sample was acquired under the same experimental conditions to address the contribution of oxygen-related defects to the emission properties observed in Ref. 28. Notably, the low-temperature O-related emission exhibits a ZPL at 590 nm, which is qualitatively in line with what reported in Ref. 8, although its prominence with respect to the baseline is non-negligible only at low temperatures. Based on this observation, we cannot rule out that the 590 nm band could be attributed to O-related impurities, although it was observed on a sample implanted with elemental Pb only in Ref. 28.
Furthermore, a band centered at 650 nm was visible at low temperatures, although it did not display a doublet and exhibited a limited prominence with respect to the baseline. The band broadened at higher wavelengths emitting in the whole 650-700 nm range. The limited spectral prominence and the difference in the spectral position (643 nm vs 650 nm) suggest that the 643 nm line observed in Pb (Refs. 28, 29) and PbO$_2$ (Ref 28 and present work) implanted samples could not be attributed to O-related impurities.

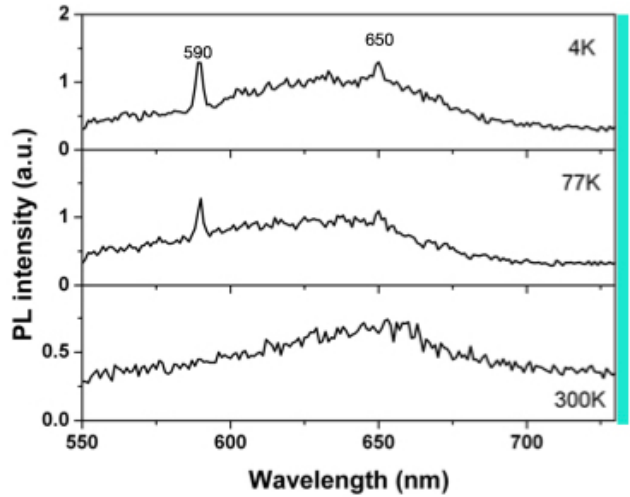

**Figure 2:** PL spectra acquired 490.5 nm optical excitation wavelengths at different temperatures (4 K, 77 K, 300 K) from the region implanted with $O^{4+}$ ions at $2\times10^{13}$ cm$^{-2}$ fluence.

*520 nm excitation*. The PL spectra (Fig. 1c) were acquired using a 50:50 beam-splitter instead of a long-pass dichroic mirror, with the purpose of allowing the collection of the spectral features at wavelengths shorter than 567 nm, corresponding to the cutoff wavelength of the available dichroic mirror. Since the above-mentioned dichroic mirror was not employed, a (514±15) nm notch filter and a long-pass filter with 530 nm cutoff wavelength were used instead, in order to filter out the laser excitation light.

The prominence of the PL peaks at 538 nm, 552 nm and 557 nm is significantly higher with respect to the other spectral features, if compared with the results obtained at shorter excitation wavelengths. The substantial intensity decrease of the 557 nm feature at increasing temperature suggests that the emission line should be attributed to PL from the Pb-related center, rather than to the first-order Raman emission of diamond (expected at 558.7 nm wavelength), more so considering that this emission line is also visible under different excitation wavelength. Therefore, in our interpretation the first-order Raman line is strongly convoluted with the PL emission feature, while being characterized by a significantly lower emission intensity. The 590 nm band is visible in the spectra, regardless of the measurement temperature. Similarly to what observed under 490.5 nm excitation, the 643 nm peak and the less intense 648 nm peak tend to convolve into a broad emission band at room temperature, although their intensities with respect to the emission doublet at 552 nm and 557 nm is weaker if compared with the PL spectra acquired at lower excitation wavelengths. The emission peak at 715 nm is also barely visible at 4 K temperature, differently from the case of 490.5 nm laser excitation.

## 3.2 Single-center-level photon emission characterization

We report on the room-temperature characterization of isolated quantum emitters resulting from ions scattered outside of the region implanted at $5\times10^{11}$ cm$^{-2}$ fluence, as reported in the confocal microscopy map[33] shown in Fig. 3a). Differently from the data reported in Ref. [29] and [28] under 450 nm and 520 nm laser excitations, respectively, the characterization was performed at room temperature under 490.5 nm excitation (2.8 mW excitation power). In this case, an additional 520 nm notch filter was used to minimize the background emission due to the first-order Raman peak of diamond (expected at 521.8 nm). We discuss the properties of a typical emitter exhibiting a narrow ZPL at 643 nm and a low intensity phonon sideband, as reported in Fig. 3b. The attribution of such spectral line to a Pb-related defect was justified by the observation of the 643 nm emission in both Ref. 27 as well as in Sect. 3.1 of the present work at the ensemble level, and in Ref. 28 from individual spots, although without a systematic characterization at the single-photon emission level.

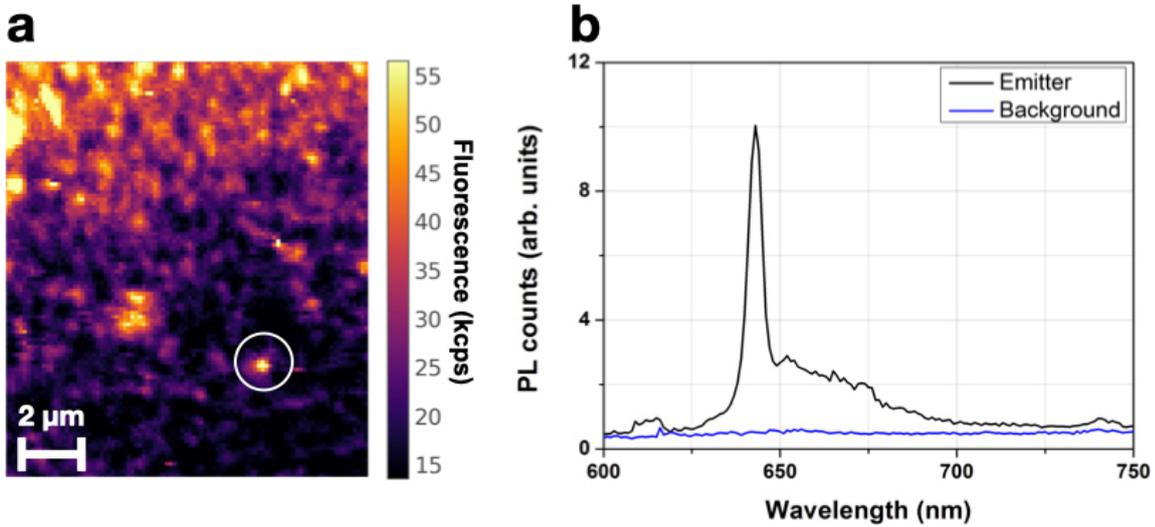

**Figure 3:** Room-temperature single-photon emission of an individual Pb-related defect under 490.5nm laser excitation. **a)** PL confocal map acquired at the edge of region implanted with $5\times10^{11}$ cm$^{-2}$ PbO$_2^-$ ions; **b)** PL emission spectrum acquired from the circled spot in **a**.

The emission properties of the center were studied by means of HBT interferometry[34,35]. Fig. 4a shows second-order auto-correlation chronograms collected under different excitation powers and background-subtracted following the procedure described in Ref. [28]. The <0.5 values observed in correspondence of zero delay time under all excitation power values is indicative of the fact that the emission originates from an individual optical center[36]. A fitting procedure of the anti-correlation dip around zero delay time was carried on the basis of the two-level system model, using the $g^{(2)}(t)=1-\exp[-\lambda \cdot t]$ functional expression, as discussed in Ref.[24]. The time constant $\tau=\lambda^{-1}$ was estimated at each excitation power. The $\lambda$ values obtained from the above-mentioned exponential fitting procedure as a function of excitation power P were linearly interpolated with the $\lambda(P) = \tau_0^{-1} + \alpha \cdot P$ function (Fig. 4b), thus allowing the estimation of the radiative lifetime of the center's excited state as $\tau_0 = (4.5 \pm 0.3)$ ns. This value is compatible with the value predicted by means of DFT calculations in Ref.[30], as well as comparable with what observed for both the 552 nm emission line under 520 nm laser excitation[28] and multiple emission lines (520, 552, 555, 575, 648 nm, see **Table 1**) under 450 nm excitation[29]. The emission rate of the defect ($1.8 \times 10^5 s^{-1}$ at ~0.93 mW excitation power) was not directly comparable with the previous results under 520 nm laser excitation (i.e. >1 Mcps[28]), since the confocal microscopy setup adopted there was coupled into open air, and was thus characterized by a higher photon detection efficiency. It is worth noting that the emission line discussed in this analysis could not be clearly detected under the same temperature conditions at the ensemble level, due to the convolution of the 643 nm peak with additional spectral feature at room-temperature (Fig. 3b). This observation suggests that the ensemble emission is composed of different types of optical transitions originating from different types of lattice defects. Particularly, this observation suggests that the room-temperature band centered at 650 nm is the convolution of several different spectra of individual defects, each emitting at a specific set of wavelengths. To further qualify this point, a broader room-temperature investigation on the spectral emission properties of a set of individual luminescent spots is presented in the following section.

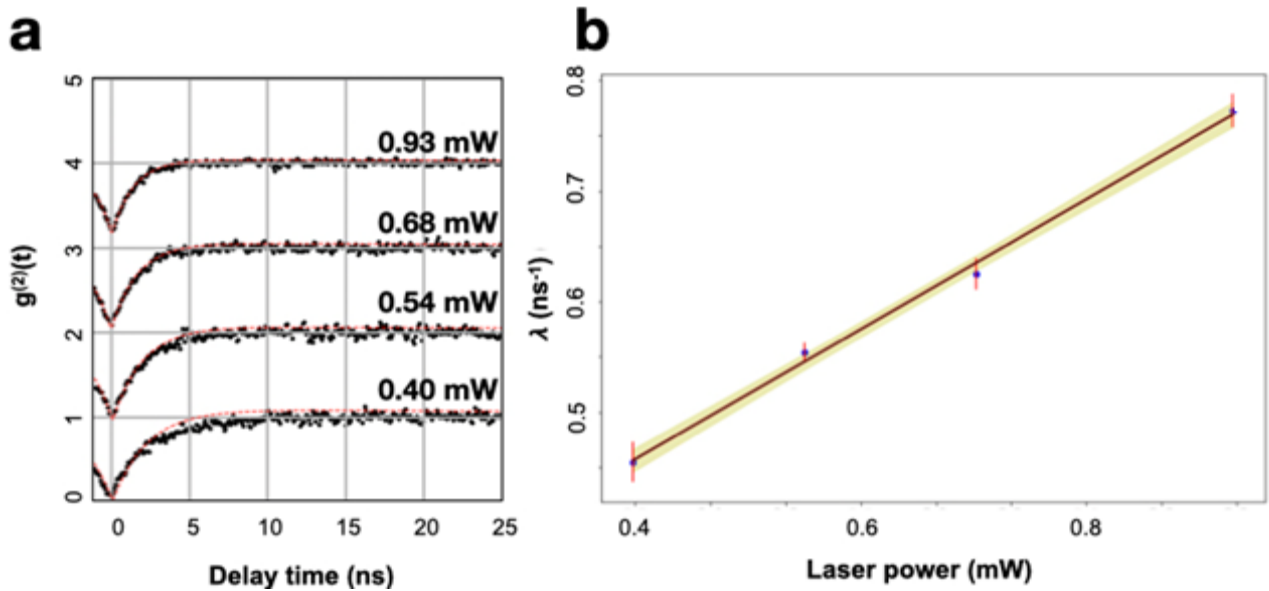

**Figure 4: a)** Background-corrected second-order auto-correlation chronograms acquired at increasing optical powers from the emitter circled in white in Fig. 3a. Each chronogram is translated by a unit step along the vertical axis for the sake of readability. **b)** Linear fit of the exponential constant value λ extracted from Fig. 4a as a function of the excitation power. The reciprocal of the intercept value defines the emission lifetime.

**3.3 Additional emission lines from isolated emission spots**

An additional study was performed to characterize individual emitters in Pb-implanted diamond. Figs. 5.a-j display a set of PL spectra acquired at room temperature under 532 nm excitation from the corresponding individual spots highlighted in the confocal map in Fig. 5k, that was acquired in correspondence of the outer edge of a $PbO_2^-$-implanted region ($5\times10^{11}$ $cm^{-2}$ fluence). For the sake of comparison, the ensemble PL spectrum acquired at the center of the implanted region under the same experimental conditions is reported in Fig. 5l. In these measurements, the use of a 550 nm long-pass filter prevented the detection of the possible occurrence of the Pb-related 552 nm line.

It is worth noting that the individual spots investigated in this analysis were not assessed by means of HBT interferometry, and thus that the observed emission spectra do not necessarily correspond to PL emission at the single-photon level.

The emission features reported in Fig. 5 can be grouped into few well-defined emission lines. Among them, the PL peaks at 557 nm and 643 nm (Fig. 5e) were already observed at the ensemble level (Fig. 5l), and can thus be safely attributed to Pb-related defects. It is worth mentioning that these two lines concurrently appear in the same spectrum, thus not ruling out that they may originate from the very same defect. Secondly, these lines do not appear concurrently with other Pb-related features discussed in the previous works, such as the 575 nm line or the 590 nm band.

An additional emission line observed in Fig. 5 is located at 554 nm (Fig. 5g). In consideration of the spectral resolution of the apparatus, this line is compatible with both the 552 nm and 557 nm lines that were attributed to Pb-related defects, and thus its attribution is not straightforward. The line could result from a lattice strain-induced shift of either the afore-mentioned PL peaks, but it is also compatible with the 555 nm emission reported in previous works from proton and neutron irradiated diamond[37,38].

In addition, a recurring set of spectral lines, not observed at the ensemble level, is visible in Fig. 5, namely: 564 nm (Figs. 5c, 5f, 5g), 567 nm (Figs. 5b, 5i, 5j), 581 nm (Fig. 5a), 605 nm (Fig. 5a), 651 nm (Figs. 5i and 5d), 655 nm (Fig. 5d), and 658 nm (Figs. 5a, 5b, 5d, 5g, 5j). There is no apparent correlation between any of these peaks, whose occurrence in the spectra reported in Figs. 5a-j is independent from all of the other observed spectral features.

This observation suggests that these emission lines are associated to different types of luminescent defects. Furthermore, the absence of such spectral features in the ensemble PL spectrum in Fig. 5l could be compatible with two possible interpretations. On one hand, the lines might be associated either to radiation-induced defects which are not directly related to Pb impurities. This attribution may be justified by previous reports of emission lines in irradiated diamond at 563 nm[37,39] and 656 nm[37]. On the other hand, some of the afore-mentioned PL peaks might be related to alternative Pb-containing defects, whose formation in high concentrations is not favored by the thermal processing adopted in the present work.

It is also worth mentioning that the 564 nm, 567 nm, 651 nm and 658 nm peaks have been observed from isolated spots at 4 K temperature (450 nm excitation wavelength) in Ref. [29]. In this work, their occurrence was significantly less frequent with respect to the main emission lines (520 nm, 525 nm, 552 nm, 575 nm, 590 nm, 638 nm) that were attributed to Pb-related defects, therefore an attribution of these emission lines was not suggested. It is therefore worth remarking that our observation is in line with the results obtained in Ref. [29], despite the adoption of a different excitation wavelength (i.e. 450 nm).

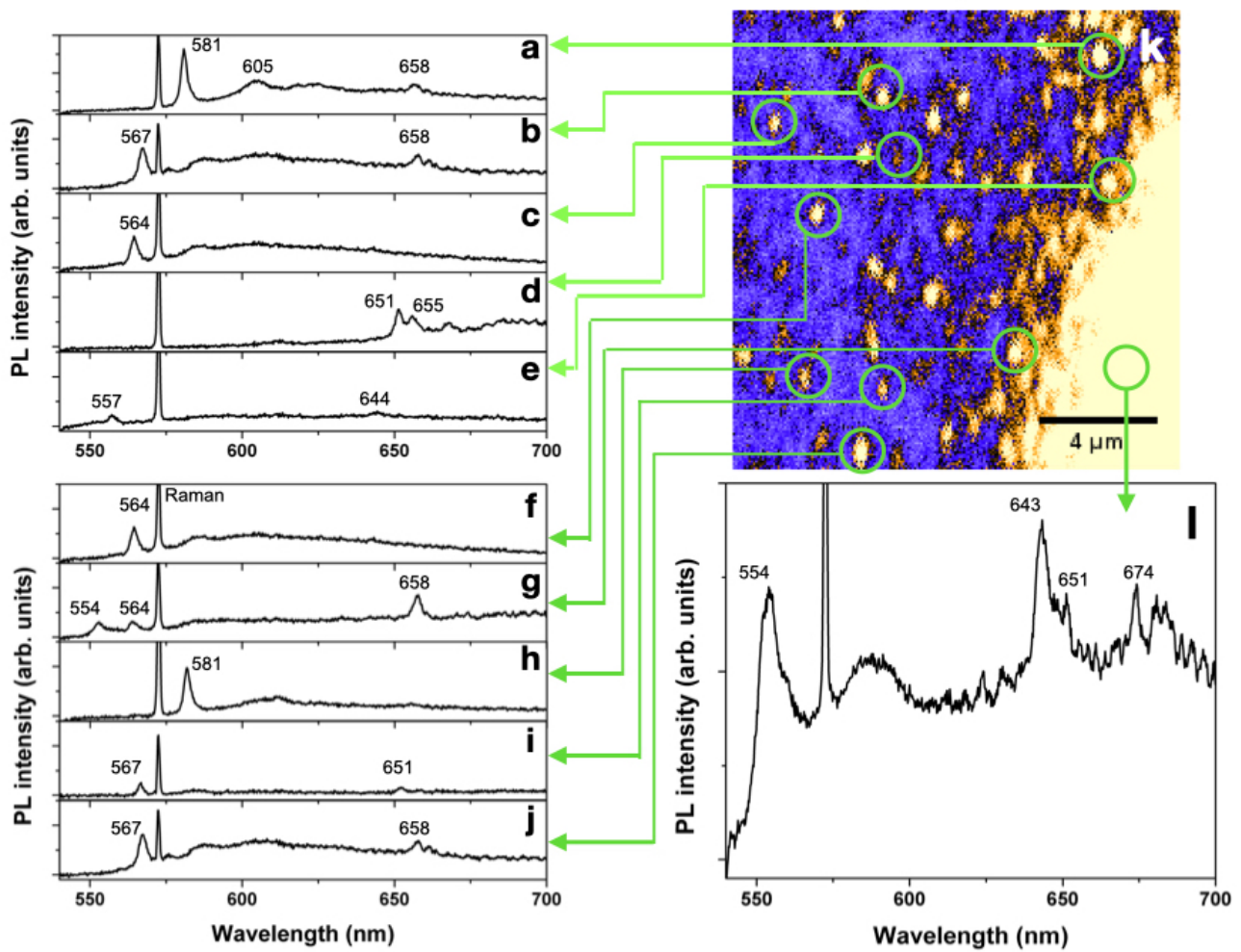

**Figure 5: a-j)** PL spectra acquired from the corresponding isolated spots circled in Fig. 5k. **k)** Confocal microscopy map acquired under 532 nm excitation wavelength at the edge of the region implanted with $5\times10^{11}$ cm$^{-2}$ PbO2$^-$. **l)** Ensemble PL spectrum acquired at the center of the implanted region under the same experimental conditions.

4. **Discussion**

In order to clarify the compatibility of the reported results with respect to previous works, the results of PL characterization at ensemble level of $PbO_2^-$-implanted diamond is discussed more extensively in this section. The reported emission lines are summarized in Table I and sorted by increasing emission wavelength, in the context of the relevant spectral features that have been reported so far. A direct comparison with the previous works[28,29] is made possible by the fact that both of them relied on the fabrication of defects by ion implantation, as well as on the same annealing temperature (i.e. 1200 °C), implying the same conditions for the formation of optical centers in diamond. The findings are also discussed in the framework of the currently available DFT theoretical predictions[27,30]. The following discussion is limited to the spectral features that were unambiguously observed at the ensemble level or directly attributed to Pb-related defects in the previous reports.

*517–520 nm.* This line was experimentally observed under 450 nm excitation at 4 K temperature in Ref. [28] from individual emission spots together with the 552 nm, 557 nm, 575 nm and 648 nm emission lines. A 2.4 eV emission is theoretically predicted to be the ZPL of the Pb-V center[29,30]. However, in our experiment we were not able to observe it under any operating condition of excitation wavelength and measurement temperature. In Fig. 6 we report the PL spectra in the 510-600 nm range acquired under different excitation wavelengths (i.e. 405 nm and 488 nm) at 4 K temperature from the region implanted with $PbO_2^-$ ions at $2\times10^{13}$ cm$^{-2}$ $PbO_2^-$ fluence. The fact that the peak was not visible under 405 nm laser excitation (Fig. 6, blue line) could be explained by assuming that the energy of the laser light might be too high to efficiently excite this emission or alternatively the occurrence of a partial ionization of the center. Furthermore, the 490.5 nm laser excitation determines the first-order 1332 cm$^{-1}$ Raman shift to be spectrally located at 524.8 nm: in this context, in this study a 490.5 nm laser excitation was adopted to allow the full deconvolution of Raman and PL spectral features across the 517-520 nm range. However, the PL spectrum collected under 409.5 nm excitation reported in Fig 6 (black line) did not exhibit any spectral feature in the region of interest, with the exception of the 525 nm Raman line. Finally, the 520 nm laser could not be adopted to investigate this line under near-resonant excitation and therefore its possible occurrence under these excitation conditions could not be assessed.

*525 nm.* This line was observed under 450 nm excitation at 4 K from an individual spot together with the 552 nm, 557 nm, 575 nm and 648nm lines[29]. However, it was not visible under 405 nm excitation in the 4–300 K temperature range and could not be assessed under 490.5nm excitation due to the overlapping first-order Raman emission at 1332 cm$^{-1}$ shift.

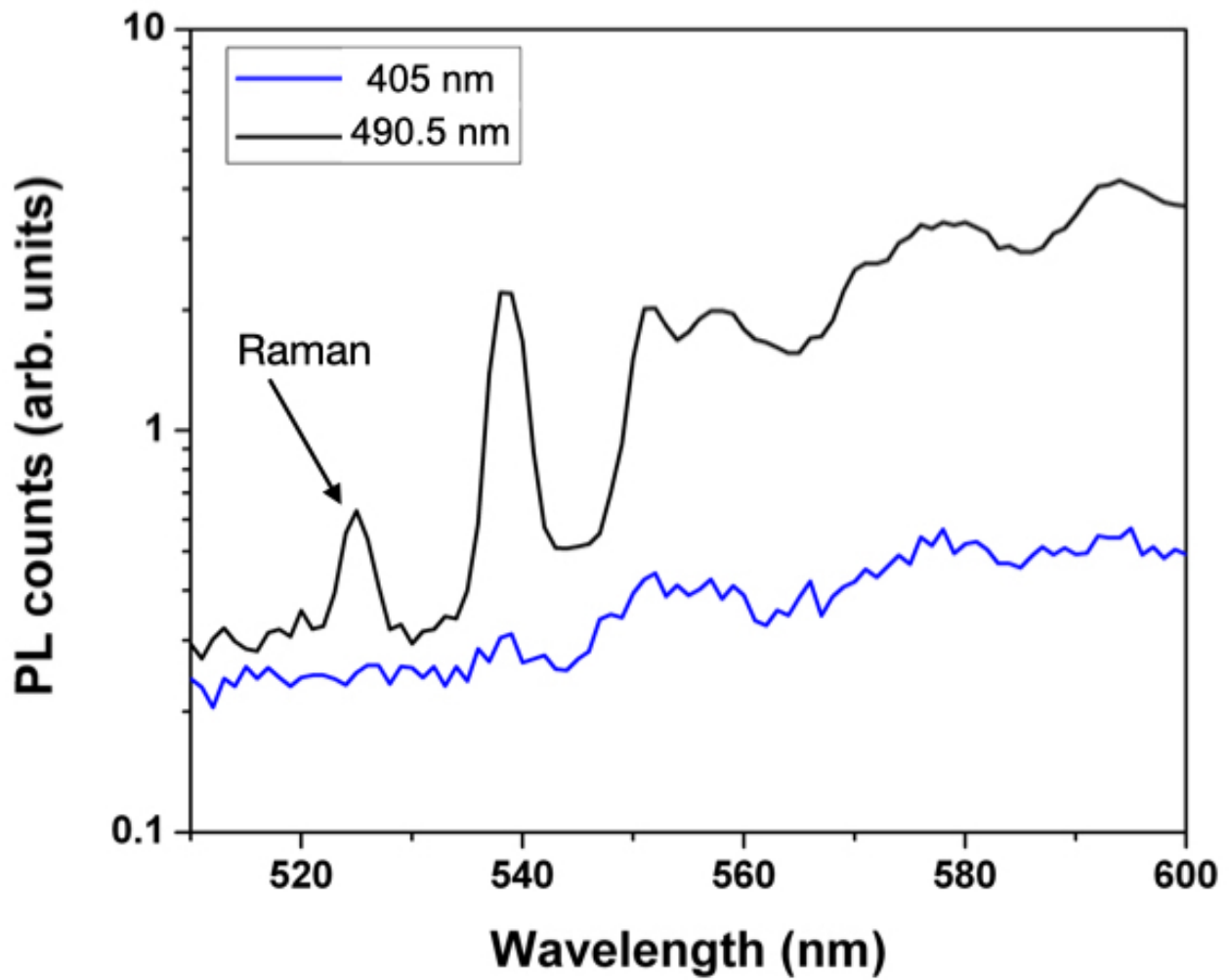

**Figure 6:** Photoluminescence spectra in the 520-600 nm range acquired from the region implanted with a fluence of $2\times10^{13}$ cm$^{-2}$ at 4 K temperature under 405 nm (blue line) and 490.5 nm excitation wavelengths (black line). The 525 nm peak corresponds to the first-order Raman shift of diamond under the 490.5 nm laser excitation.

*538 nm, 552 nm, 557 nm.* These emission peaks were observed in Ref. [28] under 514 nm and 532 nm excitations, as well as in Ref. [29] under 450 nm excitation with the exception of the 538 nm line. Their observation is confirmed in the present work under 405 nm, 490.5 nm and 520 nm excitation. The highest prominence with respect to the spectral baseline was achieved under 520 nm excitation. These emission lines were concurrently observed from isolated spots under 450 nm excitation[28], although in combination with additional lines at 525 nm and 648 nm. Also, they were observed from single-photon emitters under 532 nm excitation concurrently with the 590 nm band[28]. These results strongly suggest that all of the above-listed emission lines originate from the same type of Pb-related impurity. The 552 nm and 557 nm lines were tentatively interpreted as possible ZPL candidates for the PbV center in Ref. [28], although without the support of a comprehensive theoretical modelling.

*575 nm.* This emission line is observed in this work as a minor peak under all operating conditions. The peak is visible in the whole 4—300K temperature range under 405 nm, 450 nm, 490.5 nm, 514 nm, 520 nm and 532 nm laser excitation[28,29]. Both of the above-cited works attributed the emission to the neutrally-charged NV center, although its systematic occurrence over a number of high-quality samples characterized by a low concentration of substitutional N (~ppb) is worth being remarked.

*590 nm*. This emission band was consistently observed in all experimental reports on the Pb-related defects in the 4–300 K temperature range under 514 nm[28] and 532 nm[28,29] excitation. This work confirms its occurrence under 405 nm, 490.5nm and 520 nm excitation in all of the considered measurement temperature ranges. The band was also observed concurrently with the 538 nm, 552 nm and 557 nm lines at the single-photon emitter level, and thus tentatively interpreted as an optical-photon-mediated emission process[28]. However, the identification in this work of a 590 nm emission line in O-related ensembles suggests that this latter attribution could not be ruled out.

*639nm – 660 nm*. PL emission in this spectral region was observed in Ref. [29] at 4 K, under both 450 nm and 532 nm excitations, and it was tentatively attributed to NV centers. The observation of different lines (640 nm and 650 nm, in the 143–300 K temperature range) in Ref. [28] under 405 nm laser excitation poses questions about this attribution, since it has been previously observed that the NV⁻ emission is quenched by ionization processes at excitation energies larger than 2.6 eV[40]. In this sense, a possible, tentative attribution to a different charge state of the PbV center could not be entirely ruled out on the basis of the available data. This interpretation is mentioned in Ref. [29], where the spectral switching between the 520 nm ZPL and this set of lines could be attributed to the PbV charge state instability. It is also worth considering the recent attribution of a 639 nm PL line to interstitial-related lattice defects[41]. The plausibility of the latter interpretation is based on the large density of radiation-induced vacancies generated in diamond by the impact of an individual Pb ion.

In the present work an intense doublet was observed at the ensemble level at 643–648 nm under all experimental conditions (the highest prominence with respect to the baseline being observed under 490.5 nm), although at room temperature they were strongly convoluted into a broader emission band. The temperature-dependent results shown here offer an unambiguous link between the clear observation of the lines reported in Ref. 29 and their broadened, weak emission discussed in the Ref. 28. The fact that the 643 nm line is more intense and narrower than the 648 nm line (Fig. 1) might support the interpretation of the former as a ZPL which is associated to its first phonon replica at a larger wavelength. Notably, the ~18 meV spectral separation between the two peaks is in line with the expected value for quasi-local vibrations involving a Pb atom in the diamond lattice (i.e. ~21.5 meV[28]). Moreover, room-temperature single-photon emission was observed with a ZPL of 643 nm, as reported in Fig. 3b under 490.5 nm laser excitation. In this case the possible phonon sideband (648 nm) is broadened at room temperature over a band in the 645–680 nm range.

The ~650 nm band observed from ensembles at room temperature does not directly correspond to the spectral features of a single-photon emitter, for which the emission line is confirmed to be narrow. The band could therefore be attributed to a broadening of the ensemble emission of a set of individual emitters as a function of temperature. Conversely, a band centered at 650 nm was also observed in this work at low temperatures in O-implanted diamond, although it did not display a doublet and exhibited a limited prominence with respect to the baseline. The limited spectral prominence and the difference in the spectral position (643 nm vs 650 nm) suggest that the 643 nm line observed in Pb (Refs. 28, 29) and $PbO_2$ (Ref 28 and present work) implanted samples could not be attributed to O-related impurities.

*655-659 nm.* Some emission lines in this spectral range are reported in Ref. [29] under 450 nm excitation at 4 K temperature, however their attribution is not discussed in details. PL peaks in the 651–658 nm range were also observed at room temperature in the present work from isolated spots under 532 nm excitation, although without displaying any correlation with spectral lines related to Pb-containing defects. Our tentative attribution of these lines, due to their absence from ensemble PL spectra, is therefore of radiation-induced defects. However, it cannot be ruled out that the emission originates from a Pb-containing defect characterized by a lower formation efficiency under the thermodynamic conditions adopted both in the present study and in the previous experimental efforts[28,29].

*670 nm.* A relatively weak emission line at this wavelength is reported in Ref. [28] from room-temperature measurements carried under 405 nm laser excitation. Its observation however was not confirmed by the measurements performed in the present work under different experimental conditions (excitation wavelength, measurement temperature).

*715nm*. A peak appearing at this wavelength was reported in Ref. [28] at a 4K temperature. Particularly, it was observable under 532 nm laser excitation, while it could not be detected under a 450 nm excitation. Moreover, it was not observed at room temperature in Ref. [28] under 514 nm and 532 nm excitations. In the present work we observe that the emission line can be excited under 490.5 nm and 520 nm wavelengths at 4K and 77 K (see Fig. 1), and that its intensity significantly drops at increasing temperature until it is not visible at room temperature. This observation harmonizes the results reported Refs. [28,29], indicating that the ionization of the involved emitter occurs at excitation wavelengths across the 450–490.5 nm range, and that the line visibility is suppressed in the77–300 K temperature range. To the best of our knowledge, no apparent interpretation of this emission lines are currently available in literature.

| Emission wavelength (nm) | Excitation wavelength (nm) | Temperature | Interpretation | Notes |
|---|---|---|---|---|
| 517-520 | 450 | 4K | ZPL doublet, DFT calculations[29,30] | ° |
| 538 | 405, 490.5, 520, 532 | 4K-300K | Possible attribution as PbV ZPL[28] | |
| 552 | 450, 490.5, 520, 532 | 4K-300K | Possible attribution as PbV ZPL[28] | ° |
| 557 | 405, 450, 490.5, 520*, 532 | 4K-300K | Phonon replica of the 538/552 nm ZPL[28] | °, * |
| 565 | 450 | 4K | | |
| 575 | 450, 520, 490.5 | 4K-300K | NV[0] center ZPL[28,29] | ° |
| 590 | 405, 490.5, 520, 532 | 4K-300K | Phonon sideband of the 552 nm ZPL[28], O-related emission in PbO2 implanted diamond | |
| 639-643 | 405, 450, 490.5, 520 | 4K-300K | NV[-] center under extreme strain[29], alternative PbV charge state ZPL[28,29] | § |
| 648 | 405, 450, 490.5, 520 | 4K-300K | Phonon replica of the 643 nm ZPL | § |
| 655-659 | 450, 490.5 | 4K-300K | | |
| 715 | 490.5, 520, 532 | 4K-77K | | ** |

**Table 1:** Summary of the emission lines related to Pb ion implantation in diamond. Comments: * might be superimposed to first-order Raman under 520 nm excitation; ** extremely weak emission under 520 nm excitation; ° observed from the same emitter under 450 nm excitation; §observed from the same emitter under 490.5nm excitation.

## 5. Conclusions

We reported on the creation of Pb-related color centers in diamond by means of $PbO_2^-$ ion implantation and subsequent thermal annealing, and their optical characterization. The measurements were performed by adopting three different excitation wavelengths, namely 405 nm, 490.5 nm and 520 nm, at different temperatures in the 4–300 K range. The results of the systematic investigation of an articulated set of PL emission peaks provide a solid interpretational framework that substantially expands the understanding of this defective complex. Moreover, both of the most prominent doublets observed at 552–557 nm and 643–648 nm[28,29] were also observed at the single photon emitter level: while the former was already reported in Ref. [28] under 532 nm excitation, the latter was characterized in the present work for the first time, under 490.5nm excitation. A dedicated study on the photochromism[16] of individual defects is needed to clarify whether the striking resemblance of the two observed doublets could stem from their attributio to the ZPLs of Pb-related defects in different charge states.

The observation of several additional emission lines has also been studied under different experimental conditions. Particularly, the observation of the 715 nm line in Ref. [29] has been confirmed by our observations under 490.5 nm and 520 nm laser excitations at T≤77 K, while the emission could not be observed at room temperature.

Additionally, the ZPL at 517–520 nm observed in Ref. [29] and predicted in Ref. [30] was not observed in this work under any excitation wavelength and measurement temperature conditions.

Finally, a set of additional lines, not visible at the ensemble level, were observed from isolated spots at the edge of Pb-implanted regions. Their attribution was not conclusive and it cannot be ruled out that their lack of visibility in ensemble PL spectra might originate from a sub-optimal activation under the adopted thermal processing conditions. The high damage density introduced in the diamond lattice by the implantation of a heavy ion such as Pb might indeed require more effective processing methods to promote the formation of the PbV center, such as HPHT annealing[25], or alternative fabrication methods such as the incorporation of impurities during the crystal synthesis[26]. Future experimental studies on the defect formation based on such techniques could contribute to fully disambiguate the nature of the spectral features associated with Pb implantation in diamond.

These results clarify the potential of the PbV center for its high-impact application in quantum technologies. Besides the clear advantage offered by a bright and narrowband emitter for on-demand single-photon generation[42] or labeling in biological environments[43], the long spin coherence stemming from the predicted large ground-state splitting of the $PbV^-$ center prospects a promising tool for the practical realization of multi-qubit quantum network nodes.

**Acknowledgments**

This work was supported by the following projects: Coordinated Research Project "F11020" of the International Atomic Energy Agency (IAEA); Project "Piemonte Quantum Enabling Technologies" (PiQuET), funded by the Piemonte Region within the "Infra-P" scheme (POR-FESR 2014-2020 program of the European Union); "Departments of Excellence" (L. 232/2016), funded by the Italian Ministry of Education, University and Research (MIUR); "Ex post funding of research" of the University of Torino funded by the "Compagnia di San Paolo"; FET OPEN Pathos; INFN CSN5 Experiment "PICS4ME". The projects 17FUN01 (BeCOMe), 17FUN06 (SIQUST) 20FUN05 (SEQUME) leading to this publication have received funding from the EMPIR programme co-financed by the Participating States and from the European Union's Horizon 2020 research and innovation programme. T.L., S.P and J.M. acknowledge the support of the ASTERIQS program of the European Commission. S.D.T. and J.F. gratefully acknowledge the EU RADIATE Project (proposal 19001744) for granting transnational access to the Laboratories of the Ruđer Bošković Institute.